# Scholarly Communication[1]

Laurent Romary, INRIA & HUB

The chapter tackles the role of scholarly publication in the research process (quality, preservation) and looks at the consequences of new information technologies in the organization of the scholarly communication ecology. It will then show how new technologies have had an impact on the scholarly communication process and made it depart from the traditional publishing environment. Developments will address new editorial processes, dissemination of new content and services, as well as the development of publication archives. This last aspect will be covered on all levels (open access, scientific, technical and legal aspects). A view on the possible evolutions of the scientific publishing environment will be provided.

Keywords: digital libraries & online journals, e-print archives, communication models, communication platforms, copyright issues, open archive initiative.

---

[1] This paper has strongly benefit from close interactions with Norbert Lossau, director of the State and University Library of Göttingen.

# 1 Introduction

This chapter provides an overview of the various issues related to scientific information seen both from the point of view of the researcher, with his/her need to have access to and disseminate research results, and the research organisation, which has to define means to optimize the efficiency and visibility of the research it performs. Indeed, this chapter will provide some basic guidelines on the design of a scientific information policy that aims at the benefit of research itself.

Scientific information is considered here in the broad sense of the knowledge that a scientist acquires to carry out his/her research as well as the knowledge he/she produces, and then communicates, in the context of his/her research. This definition indeed reflects the extremely individual nature of scientific information, which relates to the capacity that researcher's results will be further used and quoted. From a wider perspective, scientific information can be defined as the knowledge that circulates within a scientific community as part of the research processes. As a consequence, defining a scientific information policy consists in optimising the scientist's information ecology, by providing the best access to existing knowledge, as well as the brightest dissemination to research productions and results.

Whereas the management of scientific information could be seen as mainly targeted to benefit the progress of science, there are some additional factors, which have to be kept in mind when trying to figure out how to organise the corresponding processes. First, managing scientific information is a kind of second-order activity, which comes, from an economical and organisational point of view, in complement, sometimes even in competition, with the actual support (in staff and equipment) that could be directly provided to research. Second, scientific information, being a tangible sign of scientific activity, is the first objective element that is taken into account when assessing the research activity of an individual, a research group, or a research organisation. As a consequence, scientific information management can seldom be considered as a pure technical activity and has often to be taken at a highly political level, whenever it impacts on the strategy of research performing — but also research funding — organisations.

Because of the intrinsic complexity of the field of scientific management, we will focus along the following sections on specific aspects that may help the reader both to identify the current trend, but also to forge himself an idea on the future evolution of the domain. After characterising scientific information at large, we will thus cover successively the domains of scientific information acquisition, the complementary issue of publication repository and open access, digital edition and data management. We will end up this chapter by providing some insights on the infrastructures that are needed for developing further scientific information services and finally by outlining what we think may be the actual information based research environment of the future.

# 2 Characterizing scientific information

There are various types of scientific information and trying to cover them all in this chapter would amount to describing all scientific fields in details. Indeed, each speciality uses and generates its own types of information and requires, when designing a scientific information policy, distinguished attention. At this stage, we can identify two main classes of scientific information:

- *Scientific publications*, which are mainly written descriptions of ideas, methods and results seen from the point of view of a specific author or group of authors;
- *Data sets*, which are acquired through the use of experimental or observational settings, as well as data production (simulations) or gathering methods such as surveys in the social sciences.

In this last category, we would integrate all the primary sources that are typically used in the humanities, either in the form of ordinary textual documents, or as specific physical artefacts (ethnographical objects, paintings, encryptions, etc.).

It is important to note here that not all types of scientific information are actually digital. In many domains, artefacts such as archaeological objects or natural components such as seeds or bacteriological samples have to be archived in specific repositories for researchers to observe or use them within experiments. Even if digital surrogates (images, scans, genomic descriptions) are often used to provide as precise a description as possible of them, they barely replace the actual physical source. In this respect, this chapter only refers to digital materials and their management for scientific purposes.

From the point of view of content, scientific information has some peculiarities that makes it differ slightly from other (possibly published) type of information:

- The relation to the author is in itself peculiar, since the content of a scholarly paper (as well as any commentary or even data) is essentially based on a personal creation, which has to be acknowledged, each time the corresponding content is being used by another party;
- Contrary to works of arts, which would be covered by the preceding point, scientific information is also subject to evaluation by the community of research, either through the peer-review process (Suls and Martin 2009), or simply by the way a research work is cited by peer colleagues;
- Scientific results should be, at least theoretically, available in the long-term (Pilat and Fukasaku 2007), since scientific knowledge can be seen as the cumulative contributions of elementary scientific results.

The combination of the three preceding points makes scientific knowledge build a network of *trusted* contributions, where no single piece can indeed be understood without an explicit reference to surrounding works.

Seen from a systemic perspective, scholarly communication processes have a very special characteristic that makes them differ from most of the types of business we may know of, namely in the relation that the business itself has with regards to the corresponding producers and customers. Indeed, researchers play a triple role in their relation to scientific information:

- As producers, since, by definition, scientific information, and in particular, publications, is the dissemination vector of research results;
- As consumers, since in most domains, existing publications and data sets play the main role in the research process, both as testimony of the current state of art or even as direct input for the establishment of new results;
- As quality controller of the scholarly communication process, by intervening at each stages where evaluation is to take place, be it to peer review a submitted paper or to assess the work of a research entity.

This intricate commitment of scholars in the scientific information workflow is actually made sustainable because *in fine* the scholar himself is rewarded from the fact that the process is actually as efficient as possible. One has to make sure that "good" research results are made available to the community, which in turn will use — and quote — them, which will result in the scholar to gain more fame (as well as institutional recognition), and thus more facilities to carry out his own research. As stated in (Edlin and Rubinfeld 2004:130): "Authors are quite different from more traditional production inputs, since part of their compensation derives from reaching readers."

Finally, it is hardly possible to speak about scientific information without tackling, at least, some basic components of the underlying technologies. Even if our purpose here is mainly to ponder upon the workflows associated to scientific information, we will see along the coming pages the strategic orientations that must be taken in order to provide science with trustful environment. This notion of trust (applicable to scientists, institutions as well as the public at large) relies on the capacity of providing reliable digital object management environments, as well as putting these within a network of interoperable components, allowing a researcher to seamlessly access and use scientific information in the various forms that we have identified so far. As we shall see such technological environment should be pragmatically designed, so that they serve a wide community of users, from big research consortia down to individual researchers.

## 3  Acquiring scientific publication

The acquisition of scholarly content through subscriptions has been for many years the main duty of research libraries, together with the acquisition of scientific books. The journal market, seen from the point of view of the library, can be characterised by a high level of stability, due basically to two main factors: a) the attachment of scholars to their favourite information sources, in a context where conservatism is the main drive for publication and reading, and b) the natural tendency not to break paper collections within the library so that one can characterise the library through its portfolio of "journals". This overall stability has been further reflected in the subscription contracts negotiated between the libraries and the publishers (or subscription agents such as SWETS or EBSCO): from one year to another the evolution pressure on the actual portfolio is essentially based on the explicit request for new titles on the part of the researchers and the identification of discarded, or less consulted, journals on the part of the library.

This intrinsic conservatism of the subscription ecology has been one of the major factors that led to the so-called *serial crisis* in the eighties. This crisis is the result of an unfortunate dynamics[2] by which publishers, taking benefit from their exclusive situation, raised the journal prices at a pace largely exceeding inflation, and libraries, facing their own financial constraints and the impossibility to cut down their core journal portfolios, focused their subscription needs on major publishers. After a few years, this crisis led, on the part of the publishers, to a quick concentration of production means, and, on the part of the libraries, to the impossibility to actually fulfil the needs of their academic communities.

In the following section, we will make an attempt at characterizing the post-serial crisis era, starting from an analysis of the notion of big deal, and exploring some possible trends to facilitate a quick and favourable evolution of the commercial scientific publication landscape.

---

[2] McCabe (2004) also demonstrates that mergers among publishers have also been a determining factor.

## 3.1 A transitional model – *Big Deals*

In the nineties, the contractual setting for subscriptions changed dramatically, following a joint necessity for publishers to optimize negotiation efforts, and, for academic libraries, to try to compensate for the high increase in price that had characterized the preceding period. This has led to the so-called "Big Deals", which have been characterized by Edlin and Rubinfeld (2004) as follows:

"in a typical Big Deal contract a library enters into a long-term arrangement to get access to a large electronic library of journals at a substantial discount in exchange for a promise not to cut print subscriptions (the prices of which will increase over time)"

As a matter of fact, such Big Deals have quickly evolved in some countries from a rather localized setting, limited to a University for instance, to cover clusters of academic institutions that started to jointly negotiate subscription agreements with publishers. After several years of practices, we would like to provide the reader with a distanced analysis of the lessons to be learnt from such big deals and see how much they may contribute to an evolution of the publishing landscape.

As a whole, Big Deals have had quite a few positive effects on the library landscape. Without analysing the actual financial issues in details, we can analyse the benefits of Big Deals by seeing how much they contributed to provide more maturity and awareness about scientific information within academic institutions:

- Big Deals participated in the creation of real community of practice among research and education libraries, since librarians had a real opportunity to compare their needs, their relations to the scholarly communities, and also their budget capacities;
- By establishing a point of focus on the relations between a given publisher and one or several institutions, they raised the political implication of academic management, but also of scientists themselves, in the decision making process;
- They speeded up the process of moving to electronic content, since most Big Deals include a wide access to the corresponding publisher's catalogue, thus bringing up the necessity of identifying the actual needs related to online access (such as long-term accessibility and archiving, see below), and the underlying infrastructures (access portals) to be put together;
- They forced academic libraries to think in-depth on the value for money that subscription schemes actually bring to institutions, in particular considering such factors as usage (hardly considered in the paper world), or budget consolidation when Big Deals would directly impact on the capacity to establish subscription contracts with smaller publishers not involved in Big Deal negotiations.

Still, these rather organisational benefits should not hide the core problems that Big Deals have brought with them, and which can be analysed along two main lines:

- Big Deals have introduced a highly conservative view on the subscription models between libraries and publishers. Indeed, the journal-based model, coming from the printed world, and the corresponding constraint imposed by publishers that turnovers should be preserved on the basis of the existing print subscription, have locked libraries into a system which give them no margin of manoeuvre in the management of library costs in the future;
- Big Deals have also showed the weaknesses of academics in dealing with strategic negotiations and in particular the difficulty for libraries to be strong at the negotiation table. Recent examples of clashes during a Big Deal negotiation should not hide the fact that many contracts have been renewed without much improvement from one round of negotiation to another.

Considering all factors, and in particular the risk that Big Deals may contribute to fossilize the subscription landscape, it is important to see such models as transient ones and identify possible evolutions that may prevent a long-term dependency to publisher's requirements. First, libraries should not be left alone in negotiating large-scale and long-term agreements with publishers. Whereas they are essential in providing objective information about the existing needs and usage, negotiation teams should always integrate academic managers and scholars with in-depth knowledge about scientific information processes. Second, prior to any negotiation, global objectives should be set, at the benefit of scholarly work, not only in terms of financial benefits, but also in terms of additional benefits that the community may gain from a new contract. We will see in the coming sections how much in particular, the notions of archival and open access may impact on the actual subscription policy of an academic institution.

## 3.2 Towards new contractual schemes

The difficulties related to the current subscription system could easily be seen as a deadlock for most research organisations since it basically hampers any possibility to define a real strategy in the domain of scientific information at large. Indeed, by freezing part of the budget on fixed expenses, it prevents these organisations from both adapting the corresponding budget to their economical situations or their priorities, but also is a major hindrance to the design of new ways for scientific information to be seamlessly exchanged among the research communities, as would be needed for better and quicker scientific progress.

Still, even if we identify that there is a need for an in-depth evolution of the publishing environment in the scholarly world, we also ascertain that it can only be implemented through the exploration of a variety of new deals between research organisations and publishers. With this orientation in mind, we present in the following sections some possible actions that could be pursued to contribute to such an evolution. All of these have been already implemented and validated within organisations and we will try to draw prospective conclusions from the assessment that we will make for each of them.

## 3.3 Open access publishing and budget centralisation

The first move we would like to address here actually comes from the publishing sector, which has experimented in the recent years ways of offering open access content (Velterop 2003) on the basis of a payment that would not result from a subscription, but that would take place on the author's side (hence the expression of *author-pays* model). In this domain, one needs to make a clear distinction between *opt-in schemes* and *native open access journals*.

In the opt-in open access schemes, publishers (e.g. the Open Choice scheme by Springer) offer the possibility for an author to finance the full accessibility on the publisher's web site of the final version of his paper at the time of publication. Such schemes have several disadvantages. It creates a burden on the authors, who have to decide on (and finance) the online publication of their papers, one by one. It implements a double payment system in institutions which do have subscriptions to the corresponding journals, and as a consequence, prevent the institutions from getting an accurate overview of the budget they dedicate to publication material. All in all, we strongly recommend not to support such schemes and to clearly inform the scholars about their danger.

True open access journals (such as the PLoS[3] portfolio or most of the titles from BMC) offer the author-pays model whereby authors must systematically pay publication fees to the corresponding publisher, but conversely no subscription is required to get access to the content. As opposed to opt-in schemes, there is here no risk of double payment and the situation is by far clearer for institutions that may centralise a specific budget made widely available to authors (thus reducing the administrative overhead). Still, such schemes (but the reverse arguments exist for the standard subscription scheme) present the dangers that it may become even more difficult to publish for institutions or countries with reduced means, and also that there may be free-riders, i.e. entities (in particular private firms) which benefit from the publication material, without contributing at all to the corresponding cost. Still, the capacity that open access journals offer to centralize the corresponding budget and thus to monitor the actual evolutions within and across research institutions worldwide encourages us to advocate the integration of open access journals within the scientific policies of academic institutions at large.

### 3.4 Making publishers' offer fit academic needs

Not only has the move from the traditional printed journal model to online delivery changed the capacity of scientists to quickly access information, it has also deeply modified the perception that academic institutions had of the management processes one should deploy for such information. This has resulted in identifying new requirements — and indeed new services — that the publishers should fulfil in conjunction with what would normally be associated with a subscription contract.

This section explores how such requirements can be contemplated, and possibly implemented, from the point of view of the content proper, i.e. how much an institution should claim to receive precise information from publishers, in order to go towards a better and more sovereign management of its own scientific information. We will indeed explore these issues in two stages, related to a) the acquisition and management of reliable meta-data and b) the various archiving levels that one may demand in the context of subscription schemes.

It is quite straightforward to understand why it is of strategic importance for an institution to have a good overview of its scientific production and consequently why managing reliable meta-data about publications is considered as a very sensitive issue. As it provides an insight on the quantity and quality of the actual publication activity of individuals as well as of institutions, such information is at the core of reporting activities, of researchers' assessment, as well as for any strategic planning endeavour. It is also of paramount importance that such metadata be both precise and accurate, since it should allow gaining insights into such a variety of publication features as the actual domains where research is produced, or the collaboration schemes that an organisation may have with other institutions or other countries for instance.

What is meant here by precise and accurate is the capacity for a database, and consequently for an interchange format, to provide metadata information where all elementary and meaningful pieces correspond to a specific field, associated with a precise semantic and mappable onto the most relevant international standards. This has an impact at three main levels of a bibliographical representation:

---

[3] Cf. Brown et al. (2003)

- In the description of publication information where article information (title, volume, issue, pagination, DOI, publication date) and journal information (title, ISSN) should allow for both a precise identification but also accurate description of the article;
- In the representation of author data, where not only names should be finely represented (all name components should have their own fields), but also affiliations should be dealt with as precisely as possible;
- In the production of basic content related information such as keywords, domain classifications or abstracts.

Whereas the first point is rather straightforward and usually dealt with quite well in various publication platforms, the second one is often underestimated. Still, precise author data and in particular affiliation represent a key aspect for the further exploitation of publication data, in particular in for the study of collaboration patterns (Subramanyam 1983). For instance, being able to sort out papers by institutions, or to analyse the geographical patterns of co-publication is part of the patterns that one wants to be able to identify out of a bibliographical database.

In the ideal case, there would be an excellent opportunity to compile such meta-data on the basis of the information available from publishers (at least for journal papers). However, as demonstrated in particular by the technical work within the European PEER project[4], scientific publishers have no coherent framework for such data and, in fact, some have difficulty to provide precise information concerning authors and their affiliation. It occurred also that publication repositories are not that in a better shape and the PEER project is the opportunity to demonstrate how much a standard-based approach (based on the customisation possibilities of the TEI[5]) could lead to a higher degree of interoperability (Ide and Romary 2004) between platforms.

If we were to make a move towards higher interoperability across platforms, the scenario contemplated here would be that any publication repository, whether public or private that contains precise and reliable chunks of information should be in the position to deliver them to complement the information that another repository would require. Whereas such a scenario may be extremely complex to implement across any kind of platform, in particular because of the heterogeneity of formats we have identified above, it is possible to contemplate the maintenance of a clearinghouse[6], which would at least serve the network of public publication repositories.

We should note here that the precision required for exchanging meta-data between publication repositories in such scenarios as the ones addressed here has nothing to do with the very shallow formats required by current harvesting processes, which, for instance, are part of protocol such as OAI-PMH (Lagoze and Van de Sompel 2001). Indeed, we anticipate an essential evolution of publication repositories, which, seen as trusted sources of information, will be able to provide, at any level of granularity, all the description attached to their digital content (see also Groth et al. 2010).

---

[4] The meta-data available from 12 publishers, comprising some of the major companies in the field, have been mapped onto one single standardized structure, in order to be pushed to a group of publication repositories in the context of a large-scale (green) open access experiment.

[5] *Text Encoding Initiative*, www.tei-c.org

[6] The PEER Depot, implemented in the context of the PEER project is an excellent example of such a clearing house.

Beyond meta-data, it is also important to consider how much an institution can demand on the availability of the full publication content and identify how much we could globally make progress in this domain. The issue here is quite simple to state: the move from paper content to digital one has changed the relation between the customers and the content in two complementary ways:

- The move to digital content has deprived step-by-step academic libraries from their role of reference archive for scholarly content (Greenstein 2000);
- Publishers have progressively taken up the management of the archival dimension of digital content, claiming that this content is always "accessible" from their own platforms, which would offer better technical facilities as well as all guarantees for a sustainable availability (Kling and McKim 1999).

These changes have created a strong dependency towards the commercial sector, with academic institutions only lately identifying the need to define a strategy in this domain. What indeed would we have is such and such publisher would disappear, or if a major conflict arises between the academic sector and the private one, where the former would lack authority towards the latter for sake of this dependency? Even if not central such a feeling has contributed to the open access movement, when scholars (and their institutions) identified that in the digital age content could, and indeed should, be pooled together in a coordinated manner.

In order to go towards a better management of published material by research institutions themselves, there is a need to act explicitly during the negotiation stage of subscription contracts. There are indeed three levels of requirements that could be identified in the domain of long-term access to publication:

- If the data is to be solely archived by the publisher, one should minimally require a *perpetual access right* to the material subscribed at a given period. Concretely, this is to guaranty that even if a contract with the publisher is interrupted, all journal content that has been paid for (new issues and backfiles) will remain accessible for the population that had benefitted from the initial agreement;
- To guaranty a better independence from the publisher's services and secure the availability of subscribed material, one can introduce a clause allowing the institution (or the corresponding consortium) to host the content and deliver it through its own servers. Further archival copies should of course be needed to ensure continuous and long-term accessibility;
- Finally, an institution can require that all the subscribed material that also corresponds to content authored by its researchers[7] could also be uploaded in a designated publication repository, in keeping with the institution's open access policy.

These three levels actually impact on the general copyright scheme that should be associated to scholarly content. While we argue that the private sector has an essential role in providing core services, such as certification, in the scientific publishing workflow, new deals have to be set so that such services are never correlated with any kind of exclusive copyright transfer.

As can be understood, we try to argue here that academic institutions should strive towards the establishment of scientific information repositories where publication meta-data and content are pooled together within a sustainable and reliable environment. Such an orientation bears a quite natural technical side, but it should also be taken very seriously as a component of ongoing discussions with the private sector. At the highest level, and in particular in the context of national licence program, such evolutions have to be discussed with the same level of attention as budget or coverage issues.

---

[7] Depending on the agreement, the coverage can be limited to corresponding authors or (like was the case of the agreement between the Max Planck Society and Springer in 2008) extended to any co-author of the publication.

## 3.5 General perspective

As a whole, we see there is a need for a real strategic view in the way we are to handle future commercial deals with the publishing sector at large. On the one hand, we need to identify means to make our budget more depending on the real need of research organisation rather than see it be determined by external interests. On the other hand, we need to identify the actual services that we require from publishers and negotiate with them the best value for money, considering also the services that we do not actually want from them. As we shall see from the following developments on research repositories, research performing organisations may want to achieve some of the functions that were so far externalized to the private sector, when these reflect a strategic interest for them.

# 4 Scientific information and open access – the role of publication repositories

## 4.1 Defining open access again?

The open access movement has gained in the recent years enough momentum and fame that one could believe that there is hardly any need to enter too much a discussion about its scope and objectives[8]. Epitomized by the Berlin declaration[9], the open access ideas reflect a tension between two opposite sets of forces:

- The necessity for the publishing business to keep its revenue, thus leading to ever-increasing subscription budgets for the academic institutions;
- The feeling that dissemination of scientific information in the digital age should be straightforward (i.e. without intermediaries) and fundamentally cheap.

With a sense that there is no clear way out of this situation, the open access movement is an echo to the feeling that academics should take action and devise their own means of disseminating their research papers. The so-called *green way to open access* is thus focussing on providing technical platforms[10] and policy settings[11], for the self-archiving and free dissemination of scholarly papers. In a way, it is complementary, or even contributes, to the attempt outlined in the first part of our paper to define new collaboration schemes between the academic institutions and the commercial publishers.

Still, the green open access movement can be characterized by a quite narrow view on scholarly publishing and the angelic idea that researchers may just join the self-deposit movement by conviction or by being mandated. The situation is *de facto* more complex, and instead of making here a theoretical or comparative analysis[12] of the publication repository landscape, we would like to illustrate the various views that may exist upon a publication repository, by presenting the specific case of the *Hyper Archive en Ligne* (HAL) repository,

---

[8] See (Davis 2009) for an overview of the doxa on the subject.
[9] http://oa.mpg.de/lang/en-uk/berlin-prozess/berliner-erklarung/
[10] usually referred to as institutional repositories (cf. Lynch 2003, and the wider distance analysis in Romary and Armbruster 2010)
[11] in the form of deposit mandates (see Sale 2007)
[12] see Armbruster and Romary (2010) for such a global comparison.

which, across the years has become the reference environment for the French academic landscape.

## 4.2 A reference case – HAL

In the mid-nineties, *Hyper Archive en Ligne* (HAL) was put together by physicists who wanted to implement a mirror archive to the already long-standing *ArXiv* (Ginsparg 1994). At that time, the main drive was to benefit from an independent platform that could have its own editorial policy, be independent from possible access problems to the United States and also be able to develop its own functionalities. With the support of the CNRS[13], a service unit[14] was put together and a first implementation made operational within several months. The spirit of this first environment was mainly targeted towards pre-prints (stage 1), to favour an early dissemination of research results.

Step by step, the French physicist community, as well as mathematicians, got used to deposit their papers in HAL and in the mid-nineties, the repository gained momentum when the CNRS decided to promote it within other disciplines and adopt it as its main source of information to published material by its researchers[15]. At the same period, several research organisations in France, among them INRIA and INSERM[16], decided to adopt HAL as a way to make concrete their adhesion to the Berlin declaration on Open Access[17]. This extension process led to a national agreement, signed in 2007, of most of the French research institutions, together with the national conference of French Universities to sign an agreement to work jointly on the further development and support of HAL.

As analysed in Armbruster and Romary (2010), the evolution quickly outlined above has set HAL as a publication repository that can hardly be reduced to the usual narrow concept of an institutional repository.

- It offers services to multiple research performing organisations such as CNRS, INSERM or INRIA, as well as universities. Each institution may have its specific portal through which their researchers may deposit and where all corresponding productions are visible. Still, any content is part of a single repository, thus allowing searches to be extended to the whole content;
- It bears a strong multidisciplinary mood, either because of the disciplinary nature of some of the supporting institutions, or because specific communities have set their own portal, but also more generally, because it matches the cultures of various scholarly communities by both considering pre-prints or published material as its core input;
- Is has reached a cross-*institutional* recognition which makes it be the default repository for Universities or research funding organisations[18] aiming at indentifying their research production.

---

[13] Centre National de la Recherche Scientifique (www.cnrs.fr)

[14] *Centre pour la Communication Scientifique Directe* (CCSD) – http://www.ccsd.cnrs.fr/

[15] cf. http://www2.cnrs.fr/journal/2546.htm

[16] *Institut National de la Santé et de la Recherche Médicale* (www.inserm.fr)

[17] http://oa.mpg.de/lang/en-uk/berlin-prozess/berliner-erklarung/

[18] See the news (in French) where the French *Agence Nationale de la Recherche* announces that it requires that all publications related to a project it funds should be deposited in HAL: http://www.agence-nationale-recherche.fr/magazine/actualites/detail/?tx_ttnews%5Btt_news%5D=159

All in all, and things being seen from the wider perspective of a national scientific information policy, HAL is providing a high level service, well integrated in the institutional landscape, for a very low budget through its highly centralized technical nature. Still, institutional recognition is, as we know, useless, if researchers themselves are not convinced that using the platform may bring some added value to their own work. It is thus worth observing why in such a community as computer science, and in particular within INRIA, HAL has gained such a fame and become part of the scientists' daily practice.

### 4.3 Publication repositories – the researcher's view

One immediate feedback that is received from colleagues using the HAL-INRIA portal on a regular basis is the high visibility that the corresponding papers receive immediately from most international search engines (hear "Google"). Indeed the centralized management of the archive has made it feasible since years to maintain good working relationship with the corresponding technical teams and to tune the software interfaces to optimize the visibility. From the point of view of the researcher, this is observed through the immediate high retrievability of a paper, as soon as relevant keywords for the corresponding field are typed in. This visibility can also be traced on the ranking of publication repositories produced by Webometrics (Björneborn and Peter Ingwersen 2004), which actually shows HAL and HAL INRIA as performing respectively on the platform and the institutional ranking.

The second important aspect, which is often mixed up with the traditional notion of institutional repository (cf. Armbruster and Romary 2010) is the specific institutional setting. Indeed HAL provides specific institutions (or communities) with dedicated portals which allow them to deal with the publication archives in their own way, by adapting the graphical charter, adopting their own editorial policy for content, or building up their additional tools (e.g. connection to in-house reporting mechanisms). Such views allow both the institution and the researchers to actually feel at home within the HAL platform, while benefiting from all generic features and evolutions that are developed globally.

Last but not least, the wide acceptance of the HAL Platform within INRIA can also be a great deal attributed to the important editorial support provided by the library network, which systematically reviews all entries after authors make a deposit, or at times even help author to make batch deposits when necessary. By completing missing meta-data, correcting actual descriptions (bibliographical descriptions, spell-checking of abstracts), by directly interacting with authors in border-line cases (software documentation, management of anteriority) and finally ensuring that the most appropriate tools and interface are being put higher-up on the developers' agenda, the librarians convey a feeling of stability that encourages even more the researchers to see HAL-INRIA as a trustful archive.

### 4.4 Perspective: intelligence in the platform

The experience we can gain from observing such a platform has HAL let us take a real distance with some of the most polemic (and at time conservative views[19]) on open access and anticipate on how much publication repositories can be made a tool at the service of science at large. Without anticipating on the wider vision we will outline at the end of the paper, we can try to make a short term project on the way we see publication archives provide even better services to researchers.

Indeed, and taking up the notion of ecology of publication repositories from (Romary and Armbruster 2010), we see that if we want to better integrate publication archives in the researcher's information ecology, we need to provide him with a variety of seamless services that will facilitate the deposit and further use of research articles. Such services would actually occur at various levels:

- At the deposit stage of a paper, one should relieve the depositor from typing in information that could be retrieved from other sources, or the paper itself. Typically, a good management of authors and associated affiliations, in particular in the context of usual co-authors is essential. In complement, automatic information extraction techniques[20] from textual content should allow to pre-fill most of the meta-data information required for a proper management of content in publication repositories;
- We should work towards providing researchers with more capacities at managing their *workspaces* within publication repositories. They should be allowed to keep drafts unpublished, add collections of graphics or images, or even additional documents such as slides, posters or videos, which may come as natural complements to the paper itself;
- As the dissemination level, it is essential to provide efficient tools for researchers to compile information from the repository, either as publication lists when preparing a paper or an institutional report, or to generate his web page or that of his research group.

---

[19] See for instance Harnad (2011).

In the current publication repository picture (cf. Armbruster and Romary 2010), such features are occasionally implemented, but there is a need for a concerted development that only a more centralized view on repositories may allow us to seize. Furthermore, institutions should clearly state how much they consider the issue of polling up publication data by redirecting all information gathering needs towards what is actually available there.

## 5  Dealing with research data and primary sources

In the continuity of the infrastructural work carried out for publication, the academic organisations are more and more turned into considering the importance of dealing with research data as part of their scientific information management duties. Indeed, as exemplified by the report issued to the European commission in October 2010 (Wood 2010), the management of data has become a key issue for the definition of new research funding programs and in particular for the establishment of digital research infrastructures. One of the key issues related to scientific data infrastructures is that of trust, in the sense that researchers may deposit and re-use scientific data with full confidence that these will be properly curated and preserved in the long run.

### 5.1 Characterising research data

The notion of research data is by far more fluid and heterogeneous as that of publication and shall deserve a specific attention. Contrary to publications, research data cannot be apprehended as a set of clearly identified objects. Data occur at various stage of the research process and may usually be seen at various levels of granularity, from elementary samples to large collections (or corpora). Quality is also an issue where part of the research activity is indeed dedicated to the selection and enrichment of primary data acquired from some equipment, or, as is the case in the humanities (Romary 2011), directly selected from various cultural heritage sources.

Scientific data bear many common features with publications. In essence, they require to be as neatly associated to the originated researcher or institution, as a warrant for the trustfulness of the content, but also in order to allow an adequate citation of the work. More generally, scientific data have to be, even more than publications, associated with precise metadata (in the same way as what we have for publications with bibliographical data). Indeed, whereas one can contemplate the idea of retrieving the source of a paper while looking at it, there is hardly any chance that a collection of numbers may tell anything of its format and origin. The fear with scientific data is that, by not taking the adequate time to document it, researchers may create data cemeteries, which may prevent any further reuse.

---

[20] See in particular the importance of machine learning techniques in this respect (Lopez 2009)

From a strategic point of view such issues cannot be dealt with without academic institutions having a real *data curation* strategy, which complements the issues of *data description* outlined above with that of *data selection*. As a matter of fact, it becomes more and more impossible for some scientific fields to keep all the data they produce, since these would go by far beyond the IT capacities as we know them today. To take the most prominent examples, the new Large Hadron Collider at CERN[21] is planned to generate several peta-bytes per year, which in turn forces their technical infrastructure to be organized as a cascading network of computer nodes, where just filtered-out data is percolated down to the individual researchers.

## 5.1 Pooling data together – the core of the scientific data business

Beyond the local management of research data at the production locus proper, it is important to keep in mind that a major challenge for the further progress of research in many fields is the capacity for research entities to be able to pool together data assets in order not only to avoid duplication of work, but even more to combine complementary evidence to form a wide-coverage informational basis for a given field. This trend has been initiated several decades ago by domains such as astronomy[22], which actually endowed themselves with various means to create networked or even joint databases of observational data (images, frequency range observations), but also of stellar objects and associated bibliographical references. To extend such trend in all fields of science, research communities, but also research organisations and governmental bodies should act in basically three directions:

- At the political level, there should be a strong incentive towards researchers so that they actually make the wide dissemination of data a standard component of their work. Such incentives may take the form of financial support, but also of effective recognition mechanisms for those who take actively part to this endeavour;
- From a legal point of view, communities should identify the least damaging licensing frameworks so that scientific results are of course systematically attributed to their producers, but also no hindrance is made to the combination of information which would have contradictory legal status[23];
- Finally, the compilation of data collections requires a strong technical coherence, to ensure in particular that data assets are made interoperable for joint queries and combined visualization or processing. This requires that scientific communities be involved in standardisation activities[24] to produce the appropriate requirements and guidelines for the representation of the data relevant for their field of activity.

---

[21] http://public.web.cern.ch/public/en/lhc/lhc-en.html

[22] See Heck (2003) for an overview, but also Heintz and Jaschek (1982) to get an idea of the visionary mood of the astronomers in the early 80's.

[23] When applicable, a basic creative commons CC-BY (http://creativecommons.org/licenses/by/3.0/), where the sole constraint is actually to guaranty the attribution of the work to its source may be considered as a baseline for such licences. Any additional constraints may then become a stumbling stone for the further pooling of information. For a wider discussion on this see Wilbanks and Boyle 2006.

As can be anticipated, the last point cannot be seen as a short-term endeavour. Each field has to identify its own requirements, link these with the existing international standards and make sure that the corresponding standards will evolve at the same pace as scientific discovery proper. As demonstrated in (Romary 2011) for the general area of the humanities, such standardisation efforts (Gray et al. 2005) relate to the capacity to provide conceptual models for the corresponding data, so that these models are also independent of the contingencies of IT facilities at a given time.

# 6 Towards a "scholarly workbench"[25] – a vision for the future of scientific information

In the previous sections, we scrutinized the ongoing evolutions of the scientific information processes, and delineated step by step a vision where institutions and scholars would take up the lead in making these changes facilitate the scholarly work. We saw in particular how much the categories of publication and data may deserve in the long run a similar treatment in a context where the separation between the two becomes less and less marked. In this context, we would like to end up our analysis by describing the way we would dream the researcher's workspace of the future, as a target object that we would try to reach when making strategic choices within our organisations. This *virtual research space* (or eSciSpace) would indeed apply for all types of scholarly activities from "hard" physical sciences to the study of primary sources in the humanities.

In the virtual research space scenario, we define a *research asset* as any piece of information that may be manipulated and further made public by a scholar in his working space. These may be pieces of text, data collections of all kinds, annotations to any research asset, or virtual *research folders* grouping together research assets in the course of the scholar's activity.

In his research workspace, the scholar manages all the stages of his scientific information activity, gathering initial evidence by importing data from external repositories, in the form of "publications" or as extractions of existing observations or documents provided by other scholars. Building up from this evidence, he compiles and organises his thoughts in the form of drafts or annotations directly linking to the other documents in his workspace, applying specific software to compute new features from his observations, and organize his thoughts by grouping meaningful sets of information in dedicated research folders.

## 6.1 Core services of the virtual research space

The virtual workspace outlined above resides on the availability of some basic services, which, not necessarily being specific to the scientific information domain, are mandatory to implement the target scenarios we anticipate. These essential services are the following ones:

- The researcher is automatically identified when connecting to his/her workspace, on the basis of a global authentication process. As a consequence, the various parameters of his/her research environment (affiliations, ongoing funded projects, co-workers) are available, so that the appropriate information can be attached easily to any research asset he/she manipulates;
- His/her workspace allows him/her to create research folders of any kind where he/she can compile various objects (notes, data sets, links to external sources, as well as related persons and events as, e.g., meetings[26]);
- Unique identifiers are provided for all objects in the workspace, which can serve as a stable reference for any further quotation or reuse;
- A precise versioning system is at hand, which allows him to trace the evolution of any research assets and publish any version according to his will.
- Last but not least, the scholar has full liability to select any of his research assets and give access rights to other individuals, groups of individuals or the public at large.

---

[24] See Ochsenbein et al. 2005 for an exemplary community endeavour.
[25] This expression appeared in the initial work plan of the eSciDoc project in 2005 (www.escidoc.org/) (Dreyer et al. 2007).

## 6.2 Community review and certification

In the perspective we want to defend here, the research workspace is also the locus where the validation of the research results takes place, opening up a whole range of new possibilities for a scholar to gain approval or recognition from his peers. Indeed, moving away from the highly focused peer-review process that we currently have for publications, the management of a research workspace combining written documents and data-sets, together with the capacity to provide various levels of visibility to any combination of these objects, allows one to anticipate a much broader community based validation of scholarly content. In the various possibilities we will describe below, the scholar always keeps the initiative to launch a certification process, but the capacity on his virtual workspace allows him actually receives the level of feedback he actually wants.

The baseline case is not that far from the current peer-review system, as we know it. It is based on the assumption that there exists certification entities associated to a specific topic, and which are articulated around a scientific committee. In this scenario, the scholar gathers research assets in a dedicated research folder and, when he considers the corresponding result to be worth it, he simply makes it accessible to the certification entity, asking for "peer review". If the evaluation is positive, the research asset receives a *review certificate*, which can be associated to any data feed related to it. Through this process, and according to the editorial policy — as well as the scientific recognition — of the certification entity, the research asset becomes a "peer-reviewed" publication in the literal meaning of the term.

On the opposite side of the spectrum, the scholar may not want to have an official certification process, but would like to receive a certain level of recognition of his results from a community. To this end, he can publicize a research asset and issue a targeted call for comment, which may range from selected colleagues to a wider scientific community. Such calls would transit through scholarly feeds or social networks to which the researcher subscribed. Feedbacks to the call take the form of *commentaries*, which may be posted in the commentators' research space and linked to the research asset. The collection of commentaries can in turn be used to derive a new version of his work or to assess the recognition of his scientific contribution within the community.

This generalized process of certification incorporates various attempts at exploring new types of peer review in various research communities. Initiatives like open peer review from the EGU (Pöschl 2004), or the attempt at creating data journals[27] for assessing the added value related to the compilation of reference data sets would naturally fit into this picture. Besides, the basically open nature of research workspace would facilitate the compilation of new types of evaluation metrics. For instance, usage statistics, coupled with information concerning the actual scholar, community of institution having used the research asset, would bring usage profiles by far more precise then any kind of citation metrics we have nowadays.

---

[26] Part of the necessary infrastructure would resemble services such as those offered by Calenda (http://calenda.revues.org/) the French diary of events in the humanities (Dacos et al. 2006).

## 6.3 The library continuum

The problem addressed with this vision of a virtual research workspace is to help the scientist manage the documents he uses and produces in his research, as well as making him feel like he is in charge of his own *virtual research library*, where he can easily retrieve documents, but also disseminate them in isolation or as collections to students, colleagues, or the wide public. Indeed such a vision decentralizes completely the notion of a research library, which would move from a central physical place to a delocalized space in the "cloud" (Aschenbrenner et al. 2009). Still, we may not expect that scholars would take up all the necessary work related to the curation of their research assets, and comprising such a variety of tasks as technical integrity or adequate meta-data description. They will have to be supported, and this is part of the infrastructure by new types of documentary experts (*eLibrarians*, or more commonly digital curators) that will accompany them in their information based research activity.

# 7 Infrastructures for scientific information

As a word of conclusion, we should situate the conditions under which the evolutions, as well as dreams, outlined in this paper could actually be implemented. Indeed, most of the developments contemplated here, whether related to publication or research data, and comprising what we did not develop here in domains such as bibliometry or digital editions, require long-standing involvements of the corresponding institutions in the domain of scientific information management. Over the years, such an involvement, together with the corresponding investments, has taken the form of real *eInfrastructures* at the service of science. Such infrastructures have shaped differently depending on the country or the academic setting ranging from high-level coordination directorate (at CNRS) or as a continuation of a strong library setting (as at the University of Göttingen, see (Lossau 2004)), it has also often been embodied as autonomous units dedicated to the management of scientific information (as in the Max-Planck Society with the Max Planck Digital Library[28]). All these service or decision units have shown from our own experience that their efficiency and capacity to contribute to the strong dynamics of the field rely on a series of essential factors:

- *Scientific digital libraries,* since they already have a name (Berman et al. 2003), should be close enough to the research communities they have to serve in order to understand both their needs but also their practices in the domain of scientific information; why indeed offer for instance information portals to scholars who would prefer using search engines to find their information sources?
- All the same, they should be constructed around a core of highly competent staff that masters such a wide range of skills as budget negotiation, IT development and knowledge representation. Scholars alone may not always have the adequate sense of the complexity related to certain decisions;
- Even more importantly, the scientific digital library should be closely related to the management of the corresponding academic institution, since from our experience, most strategic decisions in the domain of scientific information directly impact on the research environment of the institution;
- Finally, scientific digital libraries should cover the various fields of scientific information management. If their scope is limited to such a narrow domain as the management of journal subscriptions, they will not be in the position of identifying the adequate tactic moves across the various fields of scientific information.

---

[27] In a context where many data journals initiative are taking shape, we can only point to the quite visionary initiative from JISC: *Overlay Journal Infrastructure for Meteorological Sciences* (OJIMS)

[28] www.mpdl.mpg.de

Considering the current scientific information landscape, it seems that we are not far from having such scientific digital libraries at hand. Various attempts in this direction are taking place and at the same time, we better see the role and evolution of such elementary building block as publication repositories. The work ahead of us is now to bring all this together in a more coherent way.